\documentclass[conference]{IEEEtran}

\usepackage{amsmath}
\usepackage{comment}
\usepackage{amsfonts}
\usepackage{enumitem}
\usepackage{tablefootnote}
\usepackage{cite}
\usepackage{xcolor}

\ifCLASSINFOpdf
\usepackage[pdftex]{graphicx}

\else

\fi

\usepackage{algorithm}
\usepackage{algorithmic}

\ifCLASSOPTIONcompsoc
  \usepackage[caption=false,font=normalsize,labelfont=sf,textfont=sf]{subfig}
\else
  \usepackage[caption=false,font=footnotesize]{subfig}
\fi

\newcounter{myfootertablecounter}

\newcommand\myfootnotemark{%
	\addtocounter{footnote}{1}%
	\footnotemark[\thefootnote]%
}%

\newcommand\myfootnotetext[1]{%
	\addtocounter{myfootertablecounter}{1}
	\footnotetext[\value{myfootertablecounter}]{#1}
}

\usepackage{soul}

\newcommand{\sps}[1]{{\color{black}#1}}
\newcommand{\spsc}[2]{\hl{#1}{\color{black}[#2]}}
\newcommand{\ls}[1]{{\color{black}#1}}
\begin{document}

\title{Learning Generalized Wireless MAC Communication Protocols via Abstraction}

\author{
    \IEEEauthorblockN{
        Luciano Miuccio\IEEEauthorrefmark{1}, Salvatore Riolo\IEEEauthorrefmark{1}, Sumudu Samarakoon\IEEEauthorrefmark{2}, Daniela Panno\IEEEauthorrefmark{1}, and Mehdi Bennis\IEEEauthorrefmark{2}
    }
    \IEEEauthorblockA{\IEEEauthorrefmark{1} Department of Electrical, Electronics and Computer Engineering, University of Catania, Italy}
    \IEEEauthorblockA{\IEEEauthorrefmark{2} Centre for Wireless Communications, University of Oulu, Finland}
	emails: luciano.miuccio@phd.unict.it, \{salvatore.riolo, daniela.panno\}@unict.it, \{sumudu.samarakoon, mehdi.bennis\}@oulu.fi
}


\maketitle

\begin{abstract}
To tackle the heterogeneous requirements of beyond 5G (B5G) and future 6G wireless networks, conventional medium access control (MAC) procedures need to evolve to enable base stations (BSs) and user equipments (UEs) to automatically learn innovative  MAC protocols catering to extremely diverse services. This topic has received significant attention, and several reinforcement learning (RL) algorithms, in which BSs and UEs are cast as  agents, are available with the aim of learning a communication policy based on agents' local observations. However, current approaches are typically overfitted to the environment they are trained in, and lack robustness against unseen conditions, failing to generalize in different environments.
To overcome this problem, in this work, instead of learning a policy in the high dimensional and redundant observation space, we leverage the concept of observation abstraction (OA) rooted in extracting  useful information from the environment. This in turn allows learning communication protocols that are more robust and with much better generalization capabilities than current baselines.
To learn the abstracted information from observations, we propose an architecture based on autoencoder (AE) and imbue it  into a multi-agent proximal policy optimization (MAPPO) framework. Simulation results corroborate the effectiveness of leveraging abstraction when learning protocols by generalizing across environments, in terms of number of UEs, number of data packets to transmit, and channel conditions.


\end{abstract}

\begin{IEEEkeywords}
	6G, MARL, abstraction, generalization, protocol learning.
\end{IEEEkeywords}

\IEEEpeerreviewmaketitle

\section{Introduction}
Future beyond 5G (B5G) and 6G networks are envisioned to support heterogeneous services and applications, including mission-critical services, massive Internet of Things (mIoT), and so on. To meet these diversified requirements, new types of communication protocols tailored to specific applications are needed. \ls{
In this context, machine learning (ML) can be used to design new protocols with reduced time, effort, and cost compared to conventional methods \cite{9247527}.} 
In particular, multi-agent reinforcement learning (MARL) \cite{survey_2} methods 
enable agents to learn an optimal policy by interacting with non-stationary environments. Recent advances in deep RL and learning-to-communicate techniques (e.g.,  Differentiable Inter-Agent Learning (DIAL), and Reinforced Inter-Agent Learning (RIAL) \cite{citRIAL}) have led to the emergence of protocol learning for the physical (PHY) and media access control (MAC) layers \cite{nokia_paper_2, nokia_paper_1, protocols_3, protocols_4}. 
\ls{Among them, to the best of our knowledge, the only works that assess the problem of MAC protocol learning in both control and data planes are \cite{nokia_paper_2,nokia_paper_1}. Therein, user equipments (UEs) are cast as agents that learn from their partial observation of the global state how to deliver MAC protocol data units (PDUs) to the base station (BS) throughout the radio channel.
To generate optimal policies, the centralized
training and decentralized execution (CTDE) paradigm is adopted, where agents are trained offline using centralized information but execute in a decentralized manner.
Specifically, in \cite{nokia_paper_2}, both the BS and the UEs are cast as RL agents, and the multi-agent deep deterministic policy gradient (MADDPG) algorithm is adopted, that is, a commonly used CTDE-based actor-critic method.
In \cite{nokia_paper_1}, the BS is modeled as an expert agent adopting a predefined protocol, while the UEs are RL agents trained to learn a shared channel-access policy following the target signaling policy set by the BS. The policy is learned by exploiting a tabular Q-learning algorithm that follows the CTDE paradigm.
}
However, despite the good performances showed in the training environments, the learned protocols
(i.e., policies) fail to generalize outside of their training distribution, as showed in \cite{nokia_paper_1}.

We note that this drawback stems from the fact that agents learn their policies in the observational space \sps{that is specified for the environment} instead of learning observation representations \sps{that are invariant over multiple environments}, which enables better generalization and robustness.
The notion of abstraction is based on learning task structure and invariance across tasks, while filtering out irrelevant information \cite{state_abs_1}. 
Leveraging abstraction when learning a communication protocol requires tackling two questions: i) how many abstracted
observations agents need for optimal decision-making? and ii) how do agents choose their  expert policy (or a set thereof) to extract essential information? 

The main contribution of this work is to leverage abstraction to learn new wireless MAC communication protocols   with good generalization capabilities compared to state-of-the-art solutions. 
\ls{Towards this, we consider the same communication scenario presented in \cite{nokia_paper_1} and introduce the concept of observation abstraction. Specifically, we first
present a new autoencoder (AE) architecture to calculate the optimal abstraction space.
Then, we solve the cooperative MARL problem by adopting the multi-agent proximal policy optimization (MAPPO) algorithm  \cite{MAPPO_1} in the obtained abstracted observation space. We adopt MAPPO since it is one of the most promising algorithms following the CTDE paradigm for addressing cooperative MARL tasks \cite{MAPPO_1}.
Finally, the performances of the proposed solution are compared with the same MARL problem solved by adopting the MAPPO without observation abstraction and with \cite{nokia_paper_1}.
Simulation results show that the proposed approach yields policies that 
perform well not only in the training environment (as in the benchmark solutions), but also, and more importantly, in new and more complex environments that change in terms of number of PDUs, number of UEs, and channel conditions.}

The rest of the article is organized as follows. 
The system model is described in Section \ref{system} and the formalization of the cooperative MARL problem in Section \ref{marl}. The \sps{observation abstraction-based protocol learning via MARL} is detailed in Section \ref{observation}. Finally, the performance evaluation and conclusions are drawn in Section \ref{performance} and Section \ref{conclusion}, respectively.

\section {System Model}
\label{system}
\begin{figure}[!t]
	\centering
	\includegraphics[width=3.3in]{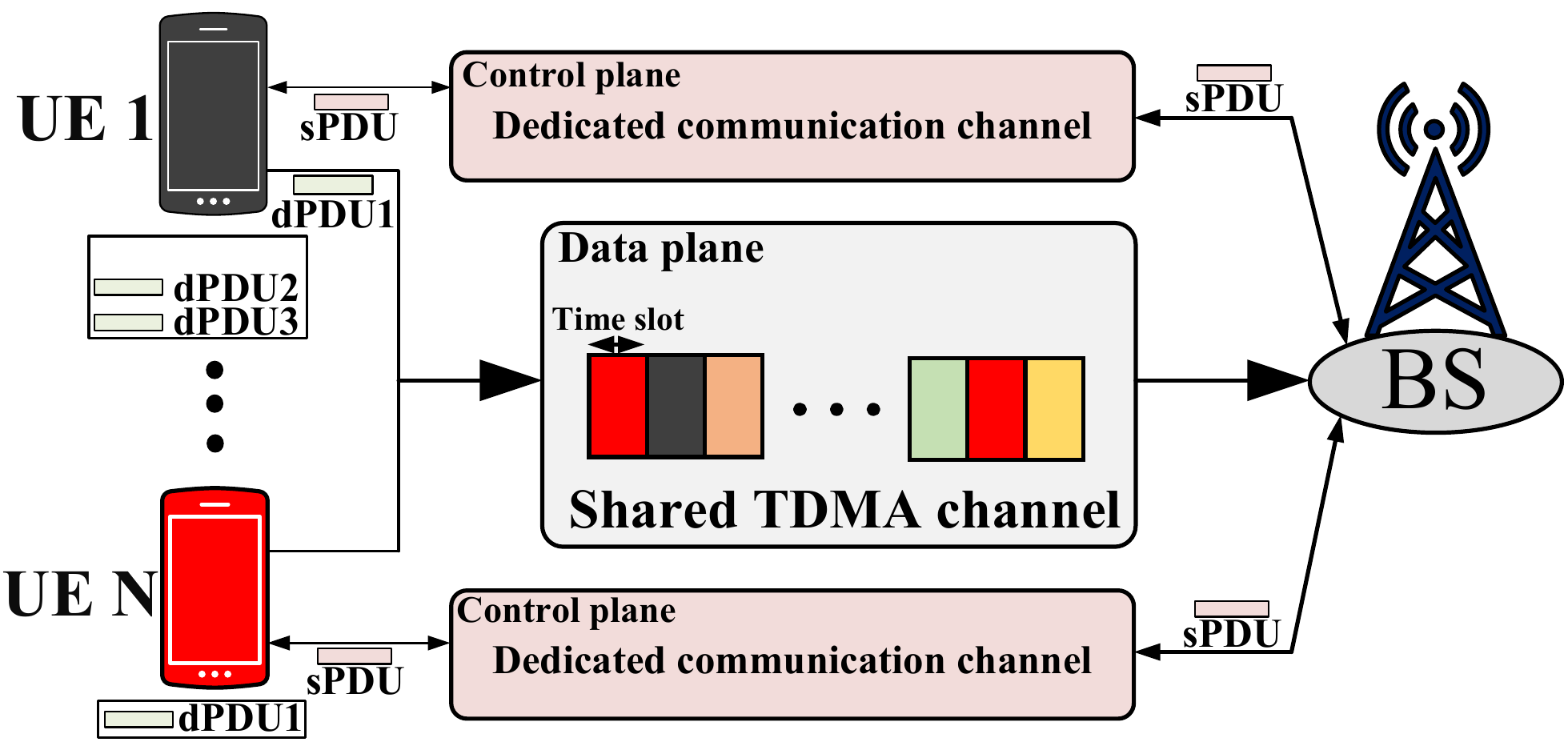}
	\caption{High-level depiction of the system model.}
	\label{system_model_fig}
\end{figure}

We consider an uplink radio network composed by \sps{a set $\mathcal{N}$ of} $N$ \sps{homogeneous} UEs and one BS, as shown in Fig. \ref{system_model_fig}. 
We consider both data plane, where the UEs transmit the 
uplink (UL) 
MAC protocol data unit (PDU) to the BS, and control plane, where UEs exchange with the BS signaling MAC PDUs. 
In the following, we denote a data PDU transmitted in the data plane with “dPDU”, a signaling PDU transmitted in the control plane with “sPDU”, and a dPDU successfully transmitted to the BS and correctly deleted from the buffer with “dPDU successfully delivered”.
The task of each UE $i \in\mathcal{N}$ is to successfully deliver $P$ dPDUs. For  sake of simplicity and to compare with \cite{nokia_paper_1}, we adopt a time division multiple access (TDMA)  channel access scheme.
In the absence of transmission errors introduced by the radio channel, a dPDU is successfully received by the BS \sps{only} if a single UE out of $N$ has transmitted its dPDU. 
\sps{If multiple UEs simultaneously transmit} 
their dPDUs, a collision occurs and the BS cannot correctly decode the received dPDUs. 


To successfully transmit their dPDUs in the same contended radio channel,  UEs (MAC learning agents) should learn an efficient MAC protocol. In the control plane, the learned MAC protocol should provide the optimal exchange of sPDUs between the UE and the BS to send the dPDU. 
Let $\mathcal{M}_{\text{UE}}$ be the set of possible uplink control plane messages, $a_{i,s}\in\mathcal{M}_{\text{UE}}$ be the signaling message sent by the $i$th UE, and $\mathcal{M}_{\text{BS}}$ be the set of 
downlink (DL) control messages. 
Moreover, we consider that the data plane transmissions are modeled as a packet erasure channel, i.e., the dPDU is correctly received with a fixed probability equal to the 
transport block error rate (TBLER). 
Conversely, 
we assume that the control channels are error-free and dedicated to each UE.

At each time slot $t$, each UE can send one sPDU to the BS in the dedicated control plane and  one dPDU in the shared data plane.
Each UE $i$ has a dPDUs storage capability, modeled as a buffer with first-in first-out (FIFO) policy, which can contain at most $P \leq Q$ dPDUs.  We denote with  $b_i^t\in\mathcal{B}=\{0,1,\dots,Q\}$ the buffer status at time $t$, and we assume $b_i^0 = P$ for all   $i \in \mathcal{N}$.
At each time slot $t$, a UE can only either transmit the first dPDU in the buffer or delete it. This means that the second dPDU in the buffer can only be considered after the first dPDU has been deleted. 

Furthermore, we assume that the BS is a MAC expert agent, i.e., it adopts a protocol that is not learned and it operates only in the DL control plane. Specifically, for each time slot $t$, the BS sends a control message $m_i^t\in\mathcal{M}_{\text{BS}}=\{0,1,2\}$ to each UE $i$. 
%
\sps{
Here, $m_i^t=2$ represents an ACK message that confirms a dPDU sent from UE $i$ has been correctly received at the BS in the previous time slot $t-1$, $m_i^t=1$ refers to a scheduling grant message to the UE $i$, and $m_i^t=0$  to indicate that no access is granted for the UE $i$.}
\ls{Clearly, the $m_i^t=2$ message can be sent to one UE at most, since the dPDU can be successfully received only if a collision was not occurred. We note that, if the UE $i$ had successfully transmitted the dPDU to the BS concurrently with the access request, then we set $m_i^t=2$. As regards $m_i^t=1$, since only one UE can be scheduled each time slot, the BS sends this message to one UE randomly chosen from the ones having transmitted the access request in $t-1$ and with $m_i^t\neq2$. }

\section {\sps{UEs-BS Interaction as a MARL Problem}} 
\label{marl}
\ls{In the considered system, the UE cannot receive information about other UEs, but makes decisions only on the basis of its observation of the global state. Moreover, UEs collaborate with one another to avoid  collisions in the shared uplink radio channel, and thus they share the same global reward. As a consequence, the protocol learning problem is cast as a cooperative and multi-agent partially observable Markov decision process (MPOMDP)  defined by}  $\langle$$\mathcal{N}$, $\mathcal{A}$, $\mathcal{S}$, $\mathcal{O}$, $\pi_{i}$, $R$, $\gamma$$\rangle$. 
\begin{itemize}
	\item $\mathcal{N}$:  set of all agents. 
	\item $\mathcal{A}$: shared action space. In this work, each agent $i\in\mathcal{N}$ shares the same action space.
	Agent $i$ performs an action ${a}_i=(a_{i,u},a_{i,s})\in\mathcal{A}$, that involves both data and control plane, where the data plane action $a_{i,u}\in\{0,1,2\}$ and $a_{i,s}\in\mathcal{M}_{\text{UE}}=\{0,1\}$. Specifically, $a_{i,u} = 1$ means that the agent transmits the first dPDU in its buffer (if any), $a_{i,u} = 2$ means it deletes the first dPDU in the buffer, and $a_{i,u} = 0$ to do nothing. For the control plane, 
	$a_{i,s}=1$ means sending an access request in order to reserve one time slot for its own transmission in the next time slot, while $a_{i,s}=0$ means do not transmit any signaling message.
	\item $\mathcal{S}$: state space of the environment.
	At time $t$, the state $s^t\in\mathcal{S}$ describes the environment by $s^t=(\textbf{b}^t,\textbf{b}^{t-1},\textbf{a}^{t-1},\textbf{m}^{t-1},\dots,\textbf{b}^{t-M},\textbf{a}^{t-M},\textbf{m}^{t-M})$, where $\textbf{b}^t=[b_1^t,b_2^t,\dots,b_N^t]$ is the vector containing all the buffer states, $\textbf{a}^t=[a_1^t,a_2^t,\dots,a_N^t]$ is the joint action vector, $\textbf{m}^t=[m_1^t,m_2^t,\dots,m_N^t]$ is the vector containing the DL control messages $m_i^t$ received by each agent $i$ from the BS, and $M$ is the memory length. 
	\item $\mathcal{O}$: set of possible observations for each agent $i$. Each agent shares the same observation space. At time $t$, each agent has a partial observation of the global state $s^t\in\mathcal{S}$, defined as $o_i^t\in\mathcal{O}$. Its observation is the tuple $o_i^t=(b_i^t,b_i^{t-1},a_i^{t-1},m_i^{t-1},\dots,b_i^{t-M},a_i^{t-M},m_i^{t-M})$. 
	\item $\pi_{i}$: policy of agent $i$, that is the probability of choosing a given action $a_i$ given its partial observation $o_i$:
	\begin{equation}
		\pi_{i}\colon\mathcal{O}\rightarrow\Delta(\mathcal{A}),
	\end{equation}
	where $\Delta$ defines a probabilistic space. Specifically, we denote with $\pi_{i}(o_i,a_i)$  the probability to take $a_i$ when observing $o_i$, and with $\pi_{i}(o_i)$ the probability distribution among all possible actions in $\mathcal{A}$ given observation $o_i$.
	\item $R\in\{-1,-\rho,+\rho\}$ is the global reward which quantifies the benefit of the joint actions performed by the agents. 
	In this regard, the agents are penalized in the following case: 
	\begin{enumerate}
		\item if there exists an agent that  deletes the dPDU without having previously transmitted it with success.
	\end{enumerate}
	Instead, the agents are positively rewarded under these conditions:
		\begin{enumerate}[resume]
		\item if there exists an agent  that  deletes its dPDU having previously transmitted it with success.
		\item If there exists an agent  that  has transmitted with success the dPDU for the first time.
	\end{enumerate}
Given these conditions, at the end of each time step $t$, each agent $i\in\mathcal{N}$ receives the same global reward $R^t$ as follows:
\begin{equation}
	R^t=\begin{cases}-\rho &\mbox{if 1) is \textbf{True}},\\
		+\rho&\mbox{ if 1) is \textbf{False} $\land$}\\
		&\mbox{  [ 2) is \textbf{True} $\lor$ 3) is \textbf{True}],}\\ -1 & \mbox{otherwise}.
	\end{cases} 
\end{equation}
	We underline that we set $R^t=-1$ when no condition is true to minimize the number of time slots. The values assigned to the reward $R^t$ follow from \cite{nokia_paper_2}.  
	However, unlike that work, where the agents are positively rewarded if condition 3) is true, we also give a positive reward if condition 2) is true, since it allows the agent to transmit, in the subsequent time steps the next packet in the buffer.
	
	\item $\gamma\in[0,1]$: discount factor, which determines the impact of  future rewards on the current decision. 
	Therefore, we define the discounted accumulated reward $G^t$ at time $t$ as:
	\begin{equation}
	G^t = R^{t} + \gamma R^{t+1} + \gamma^2 R^{t+2} + \cdots \\
	= \sum_{k=0}^{\infty} \gamma^k R^{t+k}.
	\end{equation}

%
%
\end{itemize}

%
%
%
\sps{
Due to the homogeneous nature of UEs, i.e., they share the same action space $\mathcal{A}$, the observation space $\mathcal{O}$, and the global reward $R$, instead of finding an optimal policy $\pi^*_i$ for each UE $i$, we learn a shared optimal policy $\pi^*$ via the parameter sharing technique \cite{parameter_sharing_1}.
}
%
To learn it, we adopt the CTDE paradigm. In general, several MARL techniques can be used, ranging from off-policy learning frameworks,
such as MADDPG \cite{nokia_paper_2}, value-based approaches (e.g., tabular Q-learning \cite{nokia_paper_1}), to on-policy algorithms such as MAPPO \cite{MAPPO_1}. Among them, in this work we adopt MAPPO, since its on-policy nature is well-suited to the task of learning new MAC protocols.

\section{Policy learning via abstraction}
\label{observation}

Typically, learning by abstraction is instrumental in reducing the size of the observation set $\mathcal{O}$ that can be large and contains redundant information. This can be done by clustering and aggregating
similar observations to form abstracted observations (AOs). 
In the context of RL, abstraction can overcome the fact that the  policy is overfitted to a set of redundant and noisy observations, which hurts the ability to  generalize. 
Concretely, if during the evaluation phase a new observation that was never encountered in the reduced training phase arises (e.g., a larger buffer dimension) the learned optimal policy will perform poorly. In this new environment, a new policy should be re-learned from scratch considering a large number of observations. In contrast, better generalization can be achieved by learning a policy in the abstracted observation space 
by finding the optimal solution  in the presence, during the evaluation phase, of one or many never-seen observations that are mapped into the abstracted observation space.

\subsection{Abstract Formulation}

\begin{figure}[!t]
	\centering
	\includegraphics[width=3.4in]{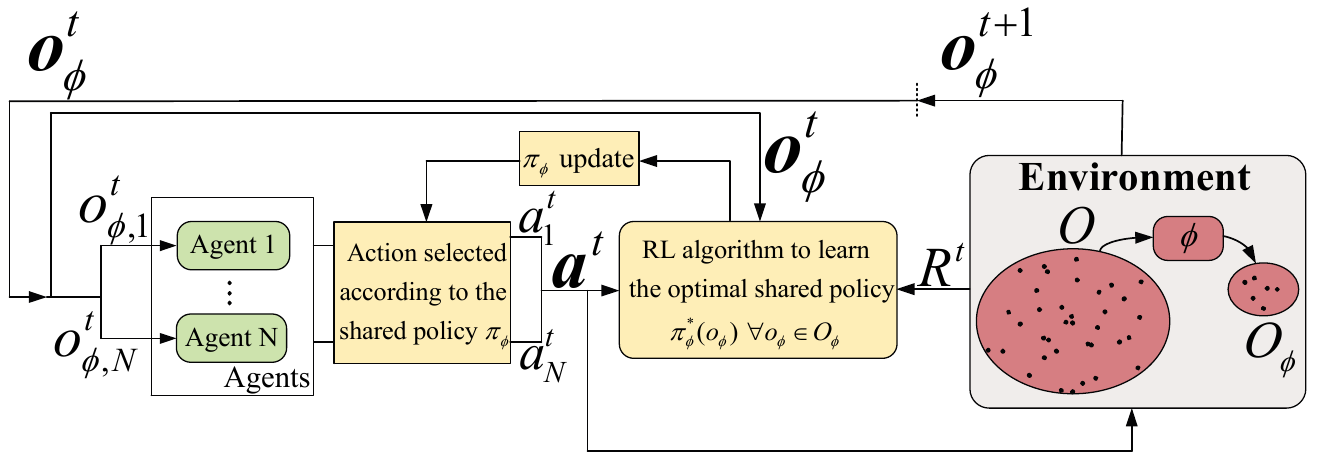}
	\caption{Proposed training procedure in the abstracted observation space.}
	\label{figura_train2}
\end{figure}

We define the abstracted MPOMDP by four components: the original MPOMDP presented in the previous Section, the observation abstraction (OA) function $\phi$, an abstraction of $\mathcal{O}$ denoted as
$\mathcal{O}_\phi$, and a shared abstracted policy operating on $\mathcal{O}_\phi$ denoted as $\pi_\phi$. Specifically, $\phi\colon\mathcal{O}\rightarrow\mathcal{O}_{\mathcal{\phi}}$
maps each observation $o_i\in\mathcal{O}$ into an abstracted observation $o_{\phi,i}\in \mathcal{O}_{\mathcal{\phi}}$. The function is injective and each agent $i$ makes use of the $\phi$ function as depicted in Fig. \ref{figura_train2}. At each time $t$, the environment provides each agent $i$ the related partial observation of  state $s^t$, denoted as $o_i^t\in \mathcal{O}$. This original observation is first passed through $\phi$ yielding the abstracted state $\phi(o_i^t)=o_{\phi,i}^t$. Then, agent $i$ uses $o_{\phi,i}^t$ to take the action $a_i^t$ according to $\pi_{\phi}$. After all agents take their actions, the environment provides one global reward $R^t$. This value, together with the vector of partial abstracted observations $\textbf{o}_\phi^t=[o_{\phi,1}^t,o_{\phi,2}^t,\dots,o_{\phi,N}^t]$ and the joint action vector $\textbf{a}^t$, are used by the RL algorithm to update $\pi_{\phi}$. As a consequence, all agents learn the abstracted shared policy
\begin{equation}
	\pi_{\phi}\colon\mathcal{O}_{\mathcal{\phi}}\rightarrow\Delta(\mathcal{A}).
\end{equation}
%
\sps{
We resort to the concept of apprenticeship learning \cite{ARGALL2009469} to find a new representation $\mathcal{O}_{\mathcal{\phi}}$ with $|\mathcal{O}_{\mathcal{\phi}}|\ll |\mathcal{O}|$
that contains the most useful information yielding efficient decision-making, i.e., $\pi_{\phi}$.
Therein,
}
learning $\pi_{\phi}$ is carried out by observing an expert demonstrator following the policy $\pi_\text{E}$ in the original observation domain. 
Hence, the goal of the observation abstraction is tantamount to compressing $\mathcal{O}$ into $\mathcal{O}_{\mathcal{\phi}}$, so that, $\mathcal{O}_{\mathcal{\phi}}$  provides  agents with an effective understanding of the environment to allow them to follow the expert policy in the abstracted space. This gives rise to an interesting trade-off between \emph{observation compression} and 
the ability of agents to follow the expert policy, expressed as a \emph{divergence} between the expect policy $\pi_\text{E}$ and the abstracted policy $\pi_{\phi}$ in the compressed space $\mathcal{O}_{\mathcal{\phi}}$. To quantify this divergence, we adopt the average Kullback-Leibler (KL) divergence:
\begin{equation}
\label{divergence}
	d\{\pi_\text{E},\pi_{\phi}\}= \underset{o\in\mathcal{O}}{\mathbb{E}}\left\{ D_{\text{KL}}(\pi_\text{E}(o)\parallel \pi_{\phi}(\phi(o))\right\},
\end{equation}
where
\begin{equation}
	D_{\text{KL}}(\pi_\text{E}(o)\parallel \pi_{\phi}(\phi(o))=\sum _{a\in {\mathcal {A}}}\pi_\text{E}(a,o)\log \left({\frac {\pi_\text{E}(a,o)}{\pi_{\phi}(a,\phi(o))}}\right).
\end{equation}

\sps{
Departing from apprenticeship learning for the single-agent MDPs that rely on a unique optimal expert policy, for the multi-agent scenario of this interest, we allow agents to adopt and exploit the information gathered from  different expert policies towards improving the robustness.
In this view, we introduce a set $\mathcal{P}_\text{E}=\{\pi^{(1)}_\text{E},\pi^{(2)}_\text{E},\dots,\pi^{(G)}_\text{E}\}$ of $G$ expert policies defined in the original observation space $\mathcal{O}$ and a set $\mathcal{P}_\phi=\{\pi^{(1)}_\phi,\pi^{(2)}_\phi,\dots,\pi^{(G)}_\phi\}$ of corresponding abstracted policies defined in the abstracted observation space $\mathcal{O}_\phi$.
}
The objective is re-defined as finding the optimal OA function with $|\mathcal{O}_\phi|\ll|\mathcal{O}|$, which minimizes the following divergence loss function:
\begin{equation}
	\label{objective}
	L_{\text{div}}=\sum_{g=1}^{G}d\left\{\pi^{(g)}_\text{E},\pi^{(g)}_\phi\right\}.
\end{equation}
%
To solve \eqref{objective}, we use AE architecture, which is composed of two deep neural networks (DNNs), namely encoder and decoder. \ls{The encoder maps the high dimensional input into a low dimensional latent representation $\textbf{z}$ of size \ls{$|\textbf{z}|$} that contains only the important information needed to represent the original input.}
The decoder reproduces the original data from $\textbf{z}$ so that the output is a representation as close as possible to the original input.
Both encoder and decoder are trained jointly to minimize the mean square error between the input and output. 

\begin{figure}[!t]
	\centering
	\includegraphics[width=3.5in]{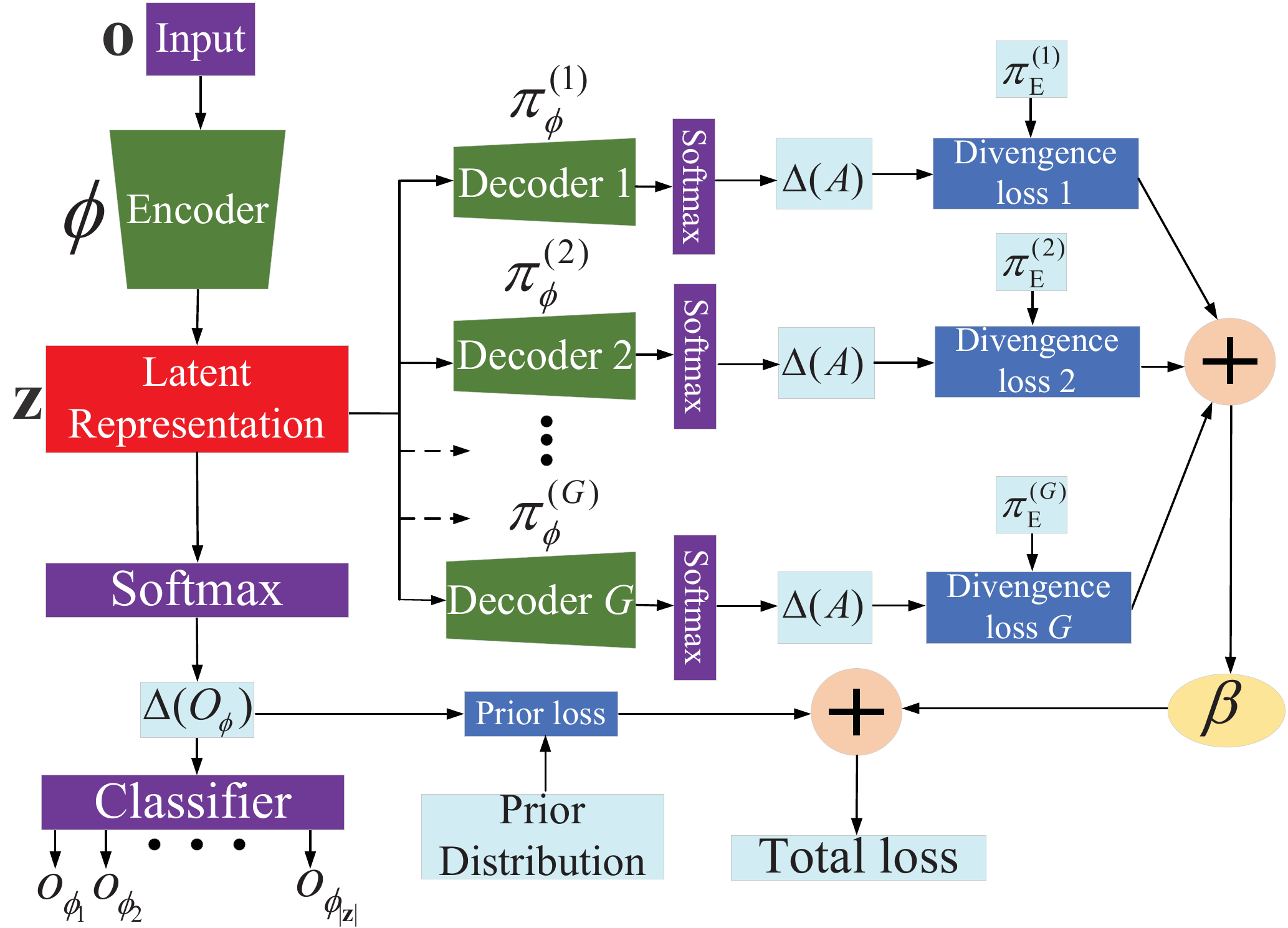}
	\caption{The proposed AE-based abstraction framework trading-off compression with value.}
	\label{Architectural_AE}
\end{figure}

Starting from the conventional AE architecture, we propose a new architecture represented in Fig. \ref{Architectural_AE}. 
Therein, the proposed AE receives the observations $o\in\mathcal{O}$ as the input.
The encoder reduces the cardinality of the input to ensure $|\mathcal{O}_{\mathcal{\phi}}|\ll |\mathcal{O}|$, rather than reducing the input dimension, as in conventional AE. 
This is realized by enforcing the proposed encoder model to act as a multi-class classifier, in which, each sample $o\in\mathcal{O}$ is  assigned to one and only one abstracted observation $o_{\phi_k}\in\mathcal{O}_\phi$ with $k\in\{1,2,\dots,|\textbf{z}|\}$ and $|\textbf{z}|\ll |\mathcal{O}|$.
Note that $|\mathcal{O}_\phi|\in\{1,2,\dots,|\textbf{z}|\}$ is held in general, since some abstracted observations $o_{\phi_k}$ may  not be assigned to any input $o$.
Therefore, the encoder represents the observation abstraction function $\phi$.
The decoder serves as an abstract policy network that maps each abstracted observation $o_{\phi_k}$ to a distribution over  action space $\mathcal{A}$ instead of reconstructing the inputs as in conventional AE.
Since we consider a set $\mathcal{P}_\text{E}$ of $G$ expert policies, we adopt $G$ decoders, where each decoder is trained to produce the $g$th abstracted policy $\pi^{(g)}_\phi$. The aim of each $g$th network is to minimize the KL divergence with respect to $\pi^{(g)}_\text{E}$ as per (\ref{objective}).
Similar to the conventional AE, both encoder and decoders are jointly trained. 
Finally, \ls{the proposed loss function is composed of the sum of two parts.}
The first one, named divergence loss, aims to achieve the goal (\ref{objective}), while the second one, named prior loss, acts as a regularization term on the latent representation  to make the distributions returned by the encoder close to a prior distribution $\textbf{p}$. 
We propose to regularize the training with a prior distribution to avoid overfitting in 
\ls{
the latent representation of the data}
so that the decoder networks can provide proper abstracted policies. For this, the regularization term is expressed as the KL divergence between the distribution at the output of the encoder and the prior $\textbf{p}$ as a uniform distribution among all the possible labels:
\begin{equation}
	\label{prior}
	L_{\text{prior}}=\underset{o\in\mathcal{O}}{\mathbb{E}}\left\{ D_{\text{KL}}(\Delta(\mathcal{O}_\phi), \textbf{p}) \right\}.
\end{equation}
The trade-off between the divergence loss and the regularization term is expressed by means of the hyper-parameter $\beta\in\mathbb{R}_{\geq0}$. The total loss is expressed as:
\begin{equation}
	\label{loss}
	L_{\text{tot}}=L_\text{{prior}}+\beta L_\text{{div}}.
\end{equation}
As $\beta\to 0$, the prior becomes more important, whereas as $\beta\to\infty$, minimizing divergence is prioritized. During training, the weights and  biases of both  encoder and  decoder models are randomly initialized and  updated  via (\ref{loss}) by using the gradient descent (GD) method for $N_\text{abs}$ episodes with the Adam optimizer and a learning rate $l_\text{abs}$. 
During evaluation, only the encoder part is adopted to provide, for each $o_i\in\mathcal{O}$, the proper label $o_{\phi_k}\in\mathcal{O}_\phi$ at the output of the classifier.

\section{Performance Evaluation}
\label{performance}
In this section, we  examine the performance of the proposed protocol learning approach leveraging abstraction, in terms of generalization to the number of dPDUs, to the TBLER, and to the number of UEs. 
\subsection{Setting}
The encoder  is a DNN composed of $3$ hidden layers, each one with 512 neurons and the rectified linear unit (ReLU) as activation function. We adopt $2$ decoders (i.e, $G=2$ expert policies), in which each decoder is a DNN with one hidden layer of 100 neurons and the ReLU as activation function. Moreover, we set $l_\text{abs}=0.00025$ and $N_\text{abs}=10000$, $\beta=1000$. The input set $\mathcal{O}$ contains all possible arrangements between the elements in $\mathcal{B}$ and the $M$-arrangements with repetition (i.e., $M$-permutations with repetition) of the elements in $\mathcal{B}$, $\mathcal{A}$, and $\mathcal{M}_\text{BS}$, with $M=1$ and $P=10$. As a consequence, $|\mathcal{O}|=2178$. As expert policies, we first adopt the conventional grant-based transmission, where the UE only transmits the dPDU following the reception of a scheduling grant, and deletes a dPDU following the reception of the ACK. A second expert policy is based on a grant-free transmission, where the UE transmits the dDPU immediately after it is available in the buffer, and deletes it after the transmission without waiting for the ACK message. 
The abstraction performance is evaluated as follows. Starting from 
$|\textbf{z}|=1$ (i.e., $|\mathcal{O}_\phi|=1$), we increased the size of $|\textbf{z}|$ with 1 unit until the loss (\ref{objective}) in the evaluation phase reached a plateau.
The optimal cardinality of $|\mathcal{O}_\phi|$ resulted equal to $8$. Therefore, we adopt it for the subsequent simulations.

For training, all the network parameters are: the number of UEs $N=2$, the number of dPDUs to transmit $P=2$, and the TBLER $= 10^{-4}$. Moreover, the training parameters together with the hyper-parameters are reported in Table \ref{table_2}. The training procedure for  MAPPO follows the approach reported in Fig. \ref{figura_train2}, both adopting as observation space $\mathcal{O}$ and the AO space $\mathcal{O}_{\phi}$. Thus, training generates two different solutions, named $M_{\mathcal{O}}$ and $M_{\mathcal{O_{\phi}}}$, respectively. 
The performance evaluation takes into account the generation of different tasks. Each evaluation task is different in terms of $P$, TBLER, and $N$. 
All the  performance results are obtained by averaging over $50$ independent simulations per configuration, and carried out in Python environment. 
%

\subsection{Benchmarks}
We compare the proposed solution, i.e., $M_{\mathcal{O_{\phi}}}$, with the $M_{\mathcal{O}}$ (no abstraction), and the approach proposed in \cite{nokia_paper_1}, named $Q_{\mathcal{O}}$.
In particular, $Q_{\mathcal{O}}$ is trained by using the same hyper-parameters as in the original paper \cite{nokia_paper_1}, and the same network parameters and reward structure of $M_{\mathcal{O_{\phi}}}$ and $M_{\mathcal{O}}$. Moreover, due to the value-based nature of the $Q_{\mathcal{O}}$ algorithm, it acts by choosing a random action when during  evaluation it encounters an observation never seen during  training.
\subsection{Results and Discussion}
We compare 
the solutions in terms of generalization to the number of dPDUs, to the TBLER, and to the number of UEs.

\begin{figure}[!t]
	\centering
	\includegraphics[width=3.2in]{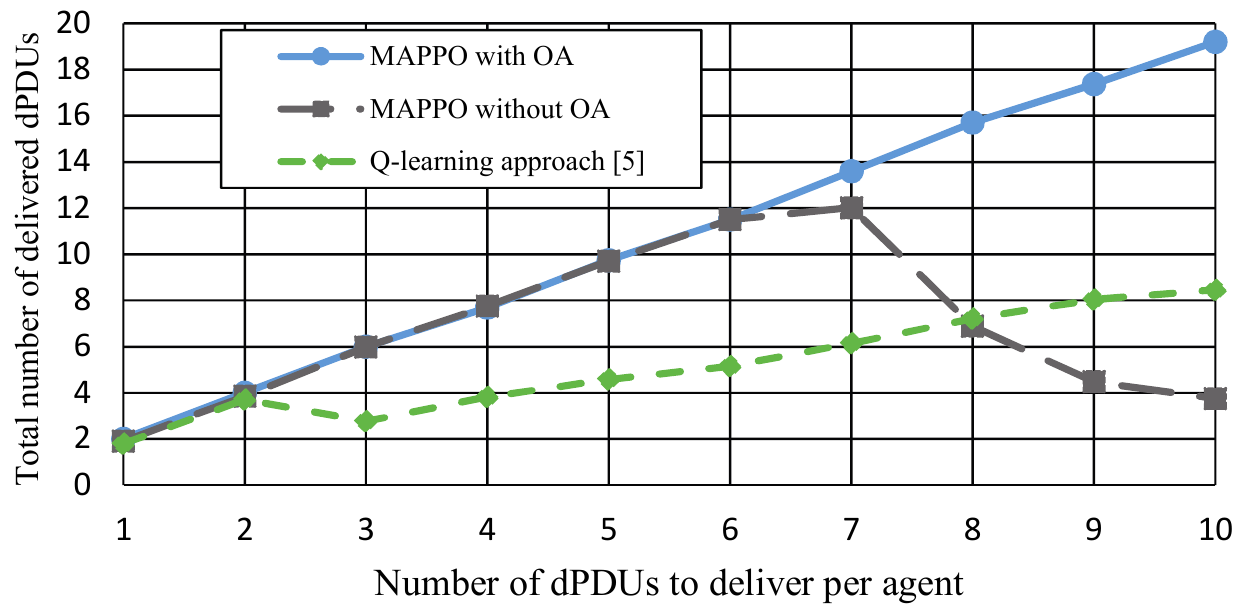}
	\caption{Average total number of successfully delivered dPDUs by the $N = 2$ agents.}
	\label{case0}
\end{figure}
\textbf{Generalization to number of dPDUs.}  Fig. \ref{case0} shows the performance in terms of total average number of successfully delivered packets when the evaluation is carried out keeping the same training parameters but $P \in [1, 2, \dots 10]$.
The results show that $Q_{\mathcal{O}}$ performs well only for the value of $P$ it was trained on, whereas its performance degrades for higher value of $Q_{\mathcal{O}}$. Conversely, $M_{\mathcal{O}}$ shows an intrinsic generalization capability, since the on-policy training induces a probabilistic behavior (trajectory) that induces a good behavior within a certain range of variation. In this case, the performances are almost perfect for $P < 7$, while in the other cases a lower performance is incurred, achieving even lower performances than the $Q_{\mathcal{O}}$ approach. Finally, $M_{\mathcal{O_{\phi}}}$ exploits the intrinsic generalization capabilities of the on-policy algorithm and jointly reduces the uncertainties related to the different observation spaces through OA, achieving almost perfect performance for all considered ranges of $P$, achieving in the most difficult configuration, i.e., $P=10$, an increment of performance of  $226.95\%$ and $512\%$ with respect to $Q_{\mathcal{O}}$ and $M_{\mathcal{O}}$.    

\begin{figure}[!htb]
    \centering
	\includegraphics[height=1.5in]{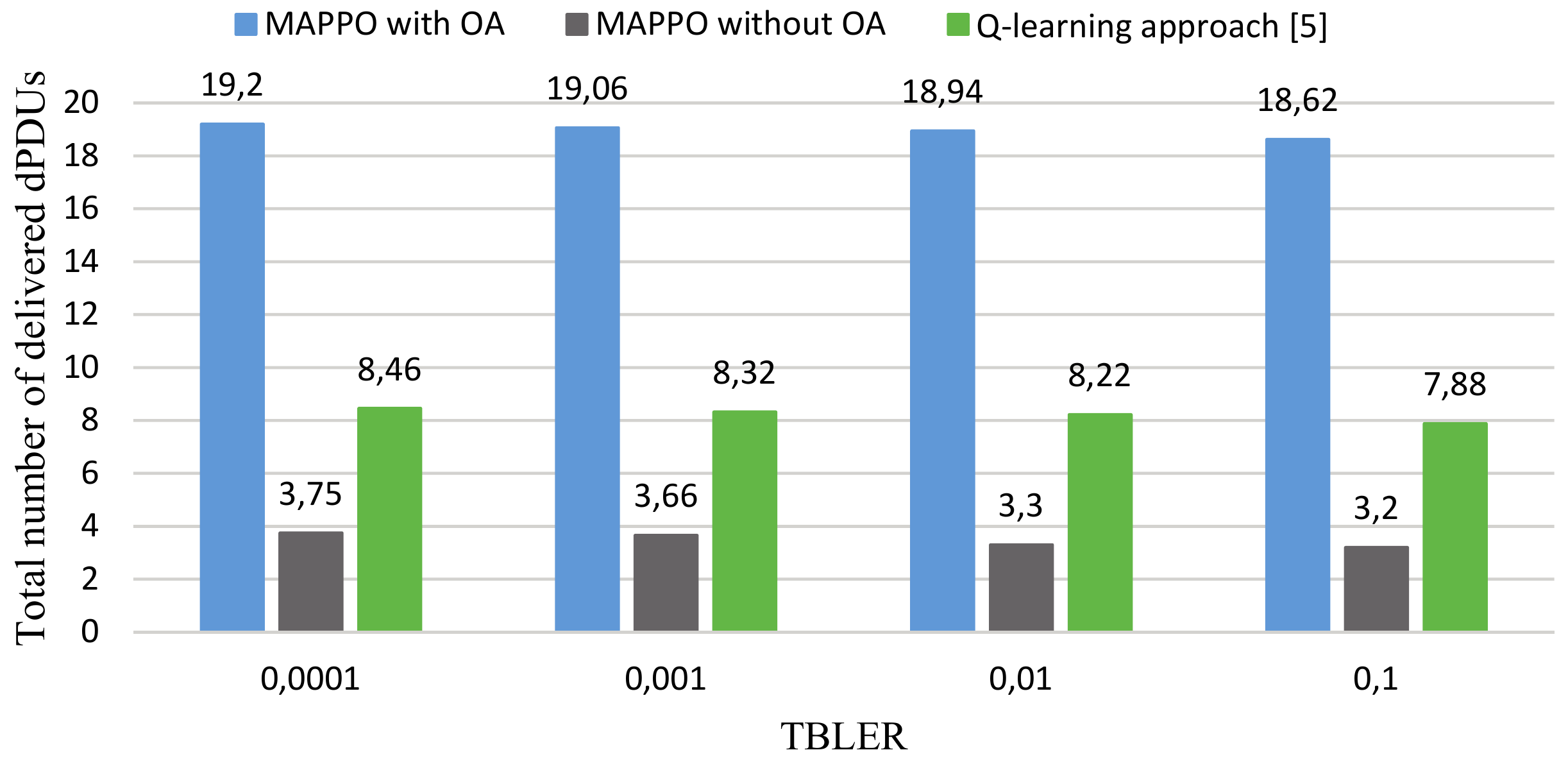} 
	\caption{Average total number of delivered dPDUs under different solutions. Training procedure with TBLER = $10^{-4}$ but the performance is evaluated with different values of TBLER.}
	\label{TBLER_fig}
\end{figure}

\begin{table}[!ht]
	\renewcommand{\arraystretch}{1.1}
	\caption{Training algorithm Parameters}
	\label{table_2}
	\centering
	\begin{tabular}{|c|c|c|}
		\hline
		\textbf{Common Parameter} & \textbf{Symbol} & \textbf{Value}\\
		\hline
		Discount factor  & $\gamma$ & 0.99 \\
		\hline
		Epsilon value & $\epsilon$ & 0.1 \\
		\hline
	 	Max. duration of episode (TTIs) & $t_{\text{max}}$ &	$300$\\
	 	\hline
	 	Reward function parameter & $\rho$& 3\\
	 	\hline
		
		\hline
		\textbf{$M_{\mathcal{O}}$ and $M_{\mathcal{O_{\phi}}}$   Parameter} & \textbf{Symbol} & \textbf{Value}\\
		\hline	
		Num. of neurons per hidden layer, evaluator & & 64\\
		\hline
		Num. of neurons per hidden layer, actor & & 64\\
		\hline
		Memory length  & $M$ & $1$ \\
		\hline		
		Learning rate &  $lr_{M}$& $10^{-3}$ \\
		\hline
		Number of training episodes & $N_{\text{tr}}$& $20$k\\
		\hline
		Act. function per layer, evaluator &&\{t, t, i\}\myfootnotemark{}\\
		\hline
		Act. function per layer, actor &&\{t, t, s\}\myfootnotemark{}\\
		\hline				
		Clipping value  & $\psi$ & $0.2$ \\
		\hline
	\end{tabular}

\end{table}
\myfootnotetext{t = tanh function, i = identity function}
\myfootnotetext{t = tanh function, s = softmax function}
\textbf{Generalization to TBLER.} To study the performances deviation related to a different value of TBLER, in Fig. \ref{TBLER_fig} we report the performance in terms of total average number of 
dPDUs successfully delivered when the evaluation is carried out with $P = 10$, $N=2$, and TBLER $\in$ [$10^{-4}$, $10^{-3}$, $10^{-2}$, $10^{-1}$]. We notice that the degradation of performance is contained for each method, and for each value of TBLER used in the evaluation. This result is obtained thanks to the technique of parameter sharing, that is used for each solution.

\begin{figure}[!htb]
        \centering
		\includegraphics[height=1.5in]{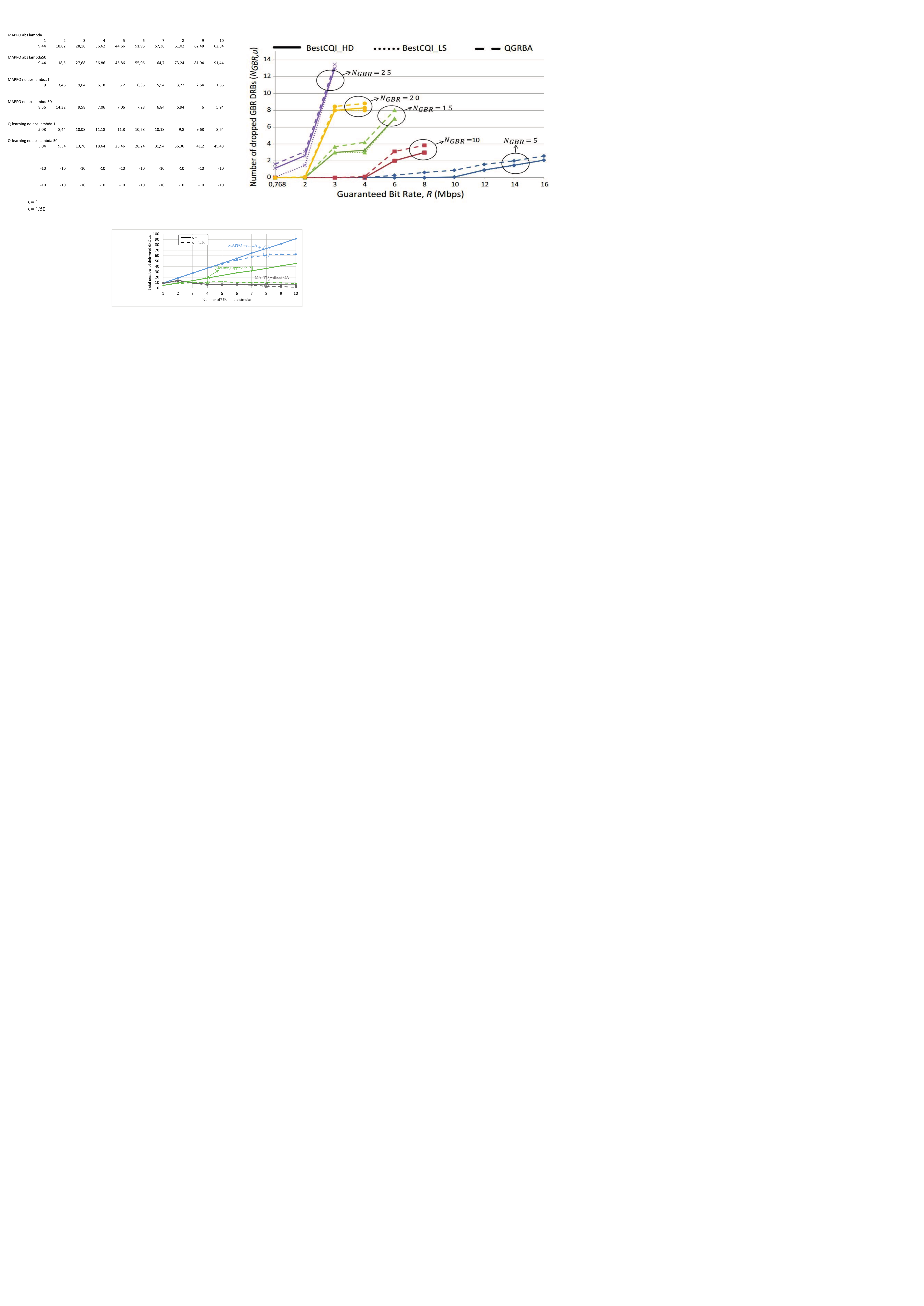} 
		\caption{Average total number of delivered dPDUs under different number of agents and Poisson arrival rate ($\lambda$). $P=10$ for each agent.}
	\label{agents}
\end{figure}

\textbf{Generalization to number of UEs.} Finally, in Fig. \ref{agents} we report generalization in terms of number of simultaneous active
UEs while setting $P=10$ and TBLER = $10^{-4}$. The UE arrivals are described by a Poisson distribution with a mean arrival rate of $\lambda$, and the simulations are carried out varying both the total number of UEs $N$ in the simulation time and $\lambda$.
The $M_{\mathcal{O_{\phi}}}$ method significantly outperforms the other solutions in any condition in terms of average total number of 
dPDUs successfully delivered. However, when the number of simultaneous active UEs is high (i.e., $\lambda = 1$ and $N \geq 7$), the performance starts to saturate.
\section{Conclusion and future work}
\label{conclusion}
In this work, we studied the problem of learning generalized MAC protocols that consider both user and control planes. To do so, we proposed a novel wireless MAC protocol learning framework for an uplink TDMA transmission scenario, based on abstraction. 
The simulation results showed that the proposed solution learns generalized MAC protocols that efficiently perform the transmission task, generalizing in terms of number of dPDUs to transmit, TBLER, and number of UEs. 
Future work will consider various extensions, such as learning a meta-protocol across various traffic classes, in addition to exploring interference-limited settings.
\section*{Acknowledgments}
This work has been partially funded by UNICT under project Piano di incentivi per la ricerca (Pia.ce.ri.) di Ateneo 2020/2022 - Linea 2D, and by the MUR under Project PON R\&I 2014-2020 Azioni IV.4 “Dottorati e contratti di ricerca su tematiche dell'innovazione”

\bibliographystyle{IEEEtran}
\bibliography{IEEEabrv,Bibliografia}

\begin{thebibliography}{10}
\providecommand{\url}[1]{#1}
\csname url@samestyle\endcsname
\providecommand{\newblock}{\relax}
\providecommand{\bibinfo}[2]{#2}
\providecommand{\BIBentrySTDinterwordspacing}{\spaceskip=0pt\relax}
\providecommand{\BIBentryALTinterwordstretchfactor}{4}
\providecommand{\BIBentryALTinterwordspacing}{\spaceskip=\fontdimen2\font plus
\BIBentryALTinterwordstretchfactor\fontdimen3\font minus
  \fontdimen4\font\relax}
\providecommand{\BIBforeignlanguage}[2]{{%
\expandafter\ifx\csname l@#1\endcsname\relax
\typeout{** WARNING: IEEEtran.bst: No hyphenation pattern has been}%
\typeout{** loaded for the language `#1'. Using the pattern for}%
\typeout{** the default language instead.}%
\else
\language=\csname l@#1\endcsname
\fi
#2}}
\providecommand{\BIBdecl}{\relax}
\BIBdecl

\bibitem{9247527}
S.~Han, T.~Xie, C.-L. I, L.~Chai, Z.~Liu, Y.~Yuan, and C.~Cui,
  ``Artificial-intelligence-enabled air interface for {6G}: Solutions,
  challenges, and standardization impacts,'' \emph{IEEE Communications
  Magazine}, vol.~58, no.~10, pp. 73--79, 2020.

\bibitem{survey_2}
K.~Arulkumaran, M.~P. Deisenroth, M.~Brundage, and A.~A. Bharath, ``Deep
  reinforcement learning: A brief survey,'' \emph{IEEE Signal Processing
  Magazine}, vol.~34, no.~6, pp. 26--38, 2017.

\bibitem{citRIAL}
J.~N. Foerster, Y.~M. Assael, N.~de~Freitas, and S.~Whiteson, ``Learning to
  communicate with deep multi-agent reinforcement learning,'' \emph{CoRR}, vol.
  abs/1605.06676, 2016.

\bibitem{nokia_paper_2}
M.~P. Mota, A.~Valcarce, J.-M. Gorce, and J.~Hoydis, ``The emergence of
  wireless {MAC} protocols with multi-agent reinforcement learning,'' in
  \emph{2021 IEEE Globecom Workshops (GC Wkshps)}, 2021, pp. 1--6.

\bibitem{nokia_paper_1}
A.~Valcarce and J.~Hoydis, ``Toward joint learning of optimal {MAC} signaling
  and wireless channel access,'' \emph{IEEE Transactions on Cognitive
  Communications and Networking}, vol.~7, no.~4, pp. 1233--1243, 2021.

\bibitem{protocols_3}
H.~B. Pasandi and T.~Nadeem, ``Towards a learning-based framework for
  self-driving design of networking protocols,'' \emph{IEEE Access}, vol.~9,
  pp. 34\,829--34\,844, 2021.

\bibitem{protocols_4}
F.~Al-Tam, N.~Correia, and J.~Rodriguez, ``Learn to schedule ({LEASCH}): A deep
  reinforcement learning approach for radio resource scheduling in the {5G MAC}
  layer,'' \emph{IEEE Access}, vol.~8, pp. 108\,088--108\,101, 2020.

\bibitem{state_abs_1}
D.~Abel, D.~Arumugam, L.~Lehnert, and M.~Littman, ``State abstractions for
  lifelong reinforcement learning,'' in \emph{Proceedings of the 35th
  International Conference on Machine Learning}, vol.~80.\hskip 1em plus 0.5em
  minus 0.4em\relax PMLR, 10--15 Jul 2018, pp. 10--19.

\bibitem{MAPPO_1}
C.~Yu, A.~Velu, E.~Vinitsky, Y.~Wang, A.~M. Bayen, and Y.~Wu, ``The surprising
  effectiveness of {MAPPO} in cooperative, multi-agent games,'' \emph{CoRR},
  vol. abs/2103.01955, 2021.

\bibitem{parameter_sharing_1}
J.~K. Terry, N.~Grammel, S.~Son, and B.~Black, ``Parameter sharing for
  heterogeneous agents in multi-agent reinforcement learning,'' \emph{CoRR},
  vol. abs/2005.13625, 2020.

\bibitem{ARGALL2009469}
B.~D. Argall, S.~Chernova, M.~Veloso, and B.~Browning, ``A survey of robot
  learning from demonstration,'' \emph{Robotics and Autonomous Systems},
  vol.~57, no.~5, pp. 469--483, 2009.

\end{thebibliography}


\end{document}